\documentclass[a4paper]{jpconf}
\usepackage{graphicx}
\usepackage{gensymb}
\usepackage{wrapfig}
\usepackage[noadjust]{cite}

\begin{document}
\title{Probing Nuclear Effects at the T2K Near Detector Using Single-Transverse Kinematic Imbalance}

\author{Stephen Dolan$^{1}$, Xianguo Lu$^{1}$, Luke Pickering$^{2}$, Tomislav Vladisavljevic$^{1}$, Alfons Weber$^{1}$ for the T2K Collaboration}

\address{$^{1}$~Denys Wilkinson Building, Keble Road, University of Oxford, Oxford, OX1~3RH, UK \\
$^{2}$ Department of Physics, Imperial College London, London, SW7 2AZ, UK}

\ead{s.dolan@physics.ox.ac.uk}

\begin{abstract}
In order to make precision measurements of neutrino oscillations using few-GeV neutrino beams a detailed understanding of nuclear effects in neutrino scattering is essential. Recent studies have revealed that single-transverse kinematic imbalance (STKI), defined in the plane transverse to an incoming neutrino beam, can act as a unique probe of these nuclear effects. This work first illustrates that an exclusive measurement of STKI at the off-axis near detector of the T2K experiment (ND280) is expected to distinguish the presence of interactions with two nucleons producing two holes (2p-2h) from alterations of the predominant underlying cross-section parameter ($M_A$ - the nucleon axial mass). Such a measurement is then demonstrated with fake data, showing substantial nuclear model separation potential.  
\end{abstract}

\vspace{-6mm}

\section{Introduction}

\begin{wrapfigure}{r}{0.5\textwidth}
\vspace{-9mm}
\includegraphics[width=0.5\columnwidth]{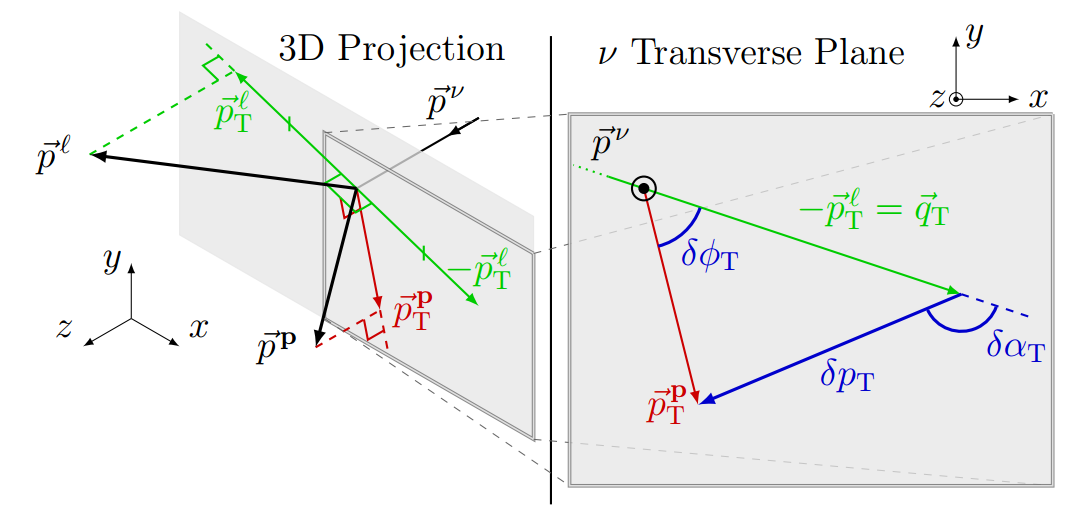}
\caption{\small A schematic ilustration of the single-transverse kinematic imbalence, $\delta \phi_\textrm{T}$, $\delta p_\textrm{T}$ and $\delta \alpha_\textrm{T}$. Here $\vec{p}^{\,\nu}$ is the incoming neutrino momentum whilst $\vec{p}^{\,\ell/\textrm{P}}$ is outgoing lepton/proton momentum. The T subscript donates projection into the plane transverse to an incoming neutrino. Taken from~\cite{luke}. }\label{fig::svtdef}
\vspace{-4mm}
\end{wrapfigure}

Accelerator-driven long baseline neutrino oscillation experiments rely on the event-by-event reconstruction of neutrino energy from interactions of few-GeV neutrino beams with nuclear targets (see, e.g.,~\cite{t2koa}). The observed final state of such interactions depends on both the interaction mode and various different nuclear effects, including Fermi motion (FM), final state interactions (FSI) and multi-nucleon correlations (np-nh), in addition to the neutrino energy. It is therefore essential to develop a detailed understanding of these nuclear effects in order to control potential bias in precision measurements of the oscillation parameters. 

It has recently been shown that single-transverse kinematic imbalance (STKI, defined in figure~\ref{fig::svtdef}), characterising imbalance between an ejected lepton and nucleon in the plane transverse to an incoming neutrino can act as excellent probes of FM and FSI in charged-current quasi-elastic (CCQE) neutrino scatters, in which a neutrino is converted to a charged lepton via the exchange of a W boson in the following reaction: $\nu_\ell n \rightarrow \ell^- p$~\cite{stv,luke,lukenpproc}. However, since nuclear effects can disguise an inelastic interaction, it is  difficult to unambiguously identify CCQE interactions in neutrino scattering experiments with heavy targets. To study `CCQE-like' scattering it is instead preferred to measure all interactions without pions in the final state (CC0$\pi$ interactions).

In this work, section~\ref{sec:stvintro} demonstrates that STKI provides an interesting probe of 2p-2h multi-nucleon correlations in realistically measurable $CC0\pi$ interactions with at least one proton in the final state ($CC0\pi+Np$ interactions). Section~\ref{sec:results} then exhibits such a measurement using fake data, extending the study presented in~\cite{steproc} to include a full error budget (to avoid bias, real data will not be analysed until the cross-section extraction machinery is fully validated). 

\vspace{-2mm}
\section{STKI and 2p-2h}
\label{sec:stvintro}

\begin{figure}[h] 
\centering
\includegraphics[width=0.95\textwidth]{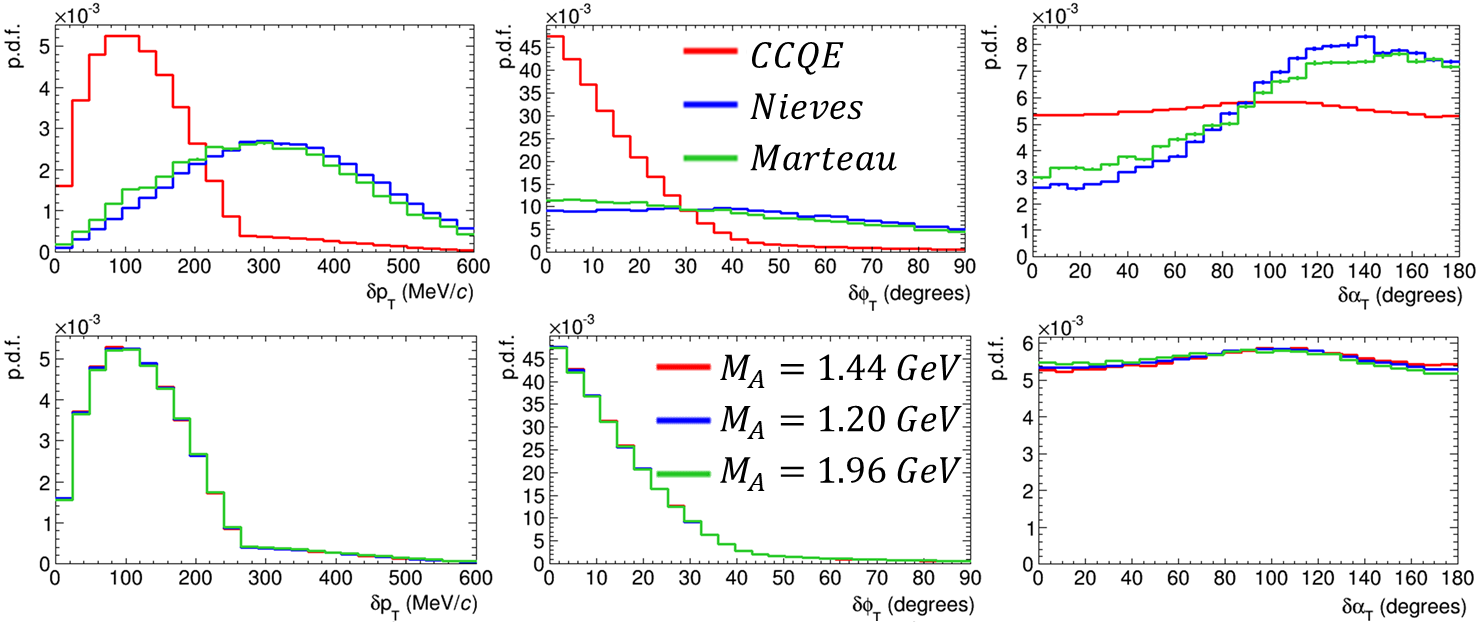}
\caption{ \small A NuWro simulation is generated, as described in section~\ref{sec:stvintro}, to investigate the ability of STKI to distinguish the scale of a 2p-2h contribution from alterations in $M_A$. The upper plots shows the shape of the 2p-2h contribution to selected interactions from two models~\cite{Nieves3, Martini1}, alongside the shape of the CCQE contribution to STKI.}
\label{fig:meccc0pi}
\vspace{-8mm}
\end{figure}

In order to effectively study 2p-2h interactions, their contribution to a measurement must be distinct from plausible variations in other models. In past cross-section measurements it has been particularly difficult to separate the scale of a 2p-2h contribution from alterations to the predominant CCQE model parameter ($M_A$ - the nucleon axial mass)~\cite{Nieves_MBComp}. Here the NuWro event generator~\cite{NuWro} is used to demonstrate that, for the implemented models, the use of STKI can allow ND280 to distinguish the effect of rescaling 2p-2h and $M_A$ in $CC0\pi+Np$ interactions. To achieve this NuWro is used to produce events in which 0.6 GeV muon-neutrinos impinge on a carbon target. Distributions of the STKI are formed from the resultant $CC0\pi+Np$ interactions which meet the the following constraints on the muon/proton momentum ($p_{\mu / \textrm{p}}$) and angle ($\theta_{\mu / \textrm{p}}$), such that they are measurable at ND280: $p_\mu>250$ MeV/\textit{c}, $p_\textrm{p}>450$ MeV/\textit{c}, $cos(\theta_{\mu})>-0.6$ and $cos(\theta_\textrm{p})>0.4$. The resultant distributions are shown in figure~\ref{fig:meccc0pi} and illustrate that a variation in $M_A$ manifests only as a normalisation shift in STKI, whilst a rescaling of 2p-2h interactions has a distinct effect on the shape, thereby inferring that a $CC0\pi+Np$ cross section at ND280 could distinguish nuclear effects from modifications to $M_A$.

\vspace{-2mm}

\section{Measuring STKI at ND280}
\label{sec:results}

ND280 is placed $2.5 \degree$ off-axis in a neutrino beam, provided by the J-PARC accelerator facility, with a peak (off-axis) energy of about $0.6$ GeV~\cite{T2KFlux}. In this analysis ND280's fully active fine grained detector (FGD1) is used as a hydrocarbon target for neutrino interactions, whilst both FGD1 and the time projection chambers (TPCs) are used for tracking. The tracker electro-magnetic calorimeter (ECal) is used as a veto for $\pi^0$ like events~\cite{T2K}. The FGD, TPCs and the ECal are situated within a 0.2T magnetic field, facilitating momentum measurements of uncontained tracks.

\begin{figure}[h] 
\centering
\includegraphics[width=1.0\textwidth]{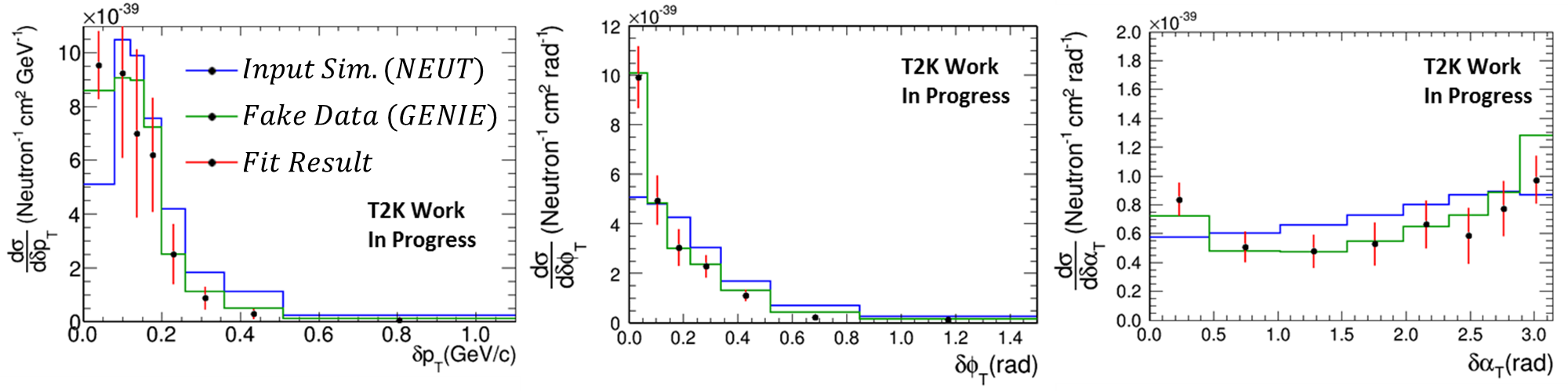}
\caption{ \small The extracted differential $CC0\pi+Np$ cross sections (with phase space constraints listed in section~\ref{sec:stvintro}) from the fake data in STKI are shown (red data points) alongside the input physics simulation from NEUT (blue line) and the fake data truth from GENIE (green line).}
\label{fig:stvxsec}
\end{figure}

The NEUT~\cite{NEUT} and GENIE~\cite{GENIE} event generators are used to act as the nominal physics simulation and produce fake data, scaled to the expected number of events in T2K runs 2-4 ($5.73 \times 10^{20}$ protons on target), respectively . The event selection and template fit based unsmearing procedure described in~\cite{steproc, cc0pi} are then used to extract the cross section of $CC0\pi+Np$ events in the fake data with same restrictions on the muon/proton kinematic phase-space as is used for the study in section~\ref{sec:stvintro}. The results are shown in figure~\ref{fig:stvxsec}, complete with statistical and systematic uncertainties representative of what can be expected for real data, compared to the GENIE fake data and NEUT input. They demonstrate the extracted cross section to be in good agreement with the fake data and to have small enough uncertainties to allow rejection of the input, indicating an effective fitting strategy and interesting model separation potential. In particular the version of GENIE used has no 2p-2h contribution which (as shown in section~\ref{sec:stvintro}) contributes to the relative deficit of events at high $\delta p_\textrm{T}$ and $\delta \phi_\textrm{T}$, whilst the large surplus of events at low $\delta p_\textrm{T}$ and $\delta \phi_\textrm{T}$ is due to GENIE's different FSI implementation. Considering this nuclear model separation potential; the shape invariance under the change of $M_A$ (demonstrated in section~\ref{sec:stvintro}) and the relative accuracy and small uncertainties predicted, it seems reasonable to expect this measurement to provide a powerful probe of nuclear effects in neutrino scattering.

\vspace{-2mm}

\vspace{-8mm}

\section{Conclusions}
\label{sec:conc}

Understanding nuclear effects in neutrino interactions is essential for making precision measurements of neutrino oscillations. It has been shown that STKI provides a unique probe of these effects, particularly through their ability to distinguish nuclear effects from alterations to the underlying cross-section model. Furthermore, it has been demonstrated that ND280 is able to exploit this interesting model separation potential by exhibiting its ability to effectively measure a $CC0\pi+Np$ differential cross section in STKI with small uncertainties relative to model differences.

\end{document}